\documentclass[aps,showpacs,amsmath,amssymb]{revtex4}
\usepackage{graphicx}

\usepackage{bm}
\usepackage{float}
\usepackage{mathtools}
\usepackage{multirow}
\usepackage{commath}
\usepackage{braket}
\usepackage{subfig}
\usepackage{bbold}
\usepackage{amsthm}
\usepackage{mathrsfs}
\usepackage[usenames]{xcolor}
\usepackage{gensymb}
\usepackage{appendix}
\usepackage{graphicx}

\usepackage[normalem]{ulem}

\DeclareMathOperator{\Tr}{Tr}

\begin{document}
\def\mean#1{\left< #1 \right>}
%\preprint{}

\title{The smallest absorption refrigerator: the thermodynamics of a system with quantum local detailed balance}

\author{Felipe Barra\footnote{fbarra@dfi.uchile.cl} and Crist\'obal Lled\'o}
\affiliation{Departamento de F\'isica, Facultad de Ciencias F\'isicas y Matem\'aticas, Universidad de Chile, Santiago, Chile}

\begin{abstract}

We study the thermodynamics of a quantum system interacting with different baths in the repeated interaction framework. 
In an appropriate limit, the evolution takes the Lindblad form and the corresponding thermodynamic quantities are determined by
the state of the full system plus baths. 
We
identify conditions under which the thermodynamics of the open system can be described only by system properties and find a
quantum local detailed balance condition with respect to an equilibrium state that may not be a Gibbs state. 
The three-qubit refrigerator introduced in \cite{linden,linden2} is an example of such a system. 
From a repeated interaction microscopic model
we derive the Lindblad equation that describes its dynamics
 and discuss its thermodynamic properties for arbitrary values of the internal coupling 
between the qubits. We find that external power (proportional to the internal coupling strength) is requiered to bring the system to its steady state, but once there, it works autonomously as discussed in \cite{linden,linden2}.

\end{abstract}
 
%\pacs{
%05.70.Ln,  %	non-equilibrium and irreversible thermodynamics 
%05.70.-a,   %    Thermodynamics
%03.65.Yz %decoherence quantum mechanics
%75.10.Pq %magnetic order spin chains
%%05.60.Gg transport process quantum
%}
\maketitle

\section{Introduction}
Quantum thermal machines can operate either in a cycle, with different numbers of strokes, or continuously~\cite[and references therein]{KosloffRev,Anders}. In either case, a proper theory of quantum thermodynamics should be able to describe the dynamics of the working medium, which we will call the system, as well as the heat fluxes coming from the environment. A continuous quantum machine normally operates in the steady-state of the equation describing the dynamics~\cite{koslov-entropy}. A great simplification is gained when this equation, which depends on all the degrees of freedom of the system and environment, can be written in closed form, describing only the degrees of freedom of the system. The most common example is the Markovian master equation in Lindblad form \cite{Lindblad, Breuer}. This simplification, however, is usually obtained at the cost of invoking approximations, which are not always well justified.

In the beginning of the studies of quantum Markovian master equations, one of the main interests was to understand how a quantum system in contact with a thermal bath dissipates energy and relaxes to equilibrium. The approximations used in that scenario, and their regime of validity, are very well understood~\cite{GKS,Davies 1, Davies 2, Spohn-Lebowitz,Spohn}. However, much of that interest has changed in the past few years due to the increasing progress of quantum thermodynamics~\cite{Anders,JGoold}. Here, many systems cannot equilibrate because they are coupled to different baths. In the standard derivation of master equations for this kind of systems the Born-Markov approximation is employed, followed by another approximation that can either make the dissipation local or global. In the local dissipation approach each bath affects a few degrees of freedom of the system while leaving the rest unchanged~\cite{Wichterich,Huelga-Plenio}. On the other hand, in the global dissipation approach each bath is responsible of transitions between energy eigenstates of the whole system~\cite{Breuer}. Since both approaches present some inconsistencies when describing the thermodynamics~\cite{Wichterich, Levy-Kosloff, Stockburger-Motz}, a lot of effort has been put recently on understanding the validity of these approaches when a system is  out of equilibrium, coupled to different baths~\cite{Huelga-Plenio, Manrique, Volovich, Purkayashta, Keeling}. It has been shown using exactly solvable models that the two approaches correctly describe the real dynamics in two different regimes of internal system couplings strength, and they complement each other \cite{Correa2017, Marti2017, Plenio2017}.

The present paper has a double purpose. The first one is to emphasize that, even though Lindblad master equations can correctly describe the dynamics under certain regimes, if a microscopic derivation is not given, one should be careful with thermodynamic statements. To this purpose, we focus on the particular case of the smallest possible quantum absorption refrigerator studied in~\cite{linden, linden2}. There, a local Lindblad master equation for a three-qubit refrigerator is postulated on phenomenological grounds, and it is argued that it can reach the ideal Carnot coefficient of performance autonomously. We microscopically derive this Lindblad equation from a repeated interaction scheme~\cite{Orszag, Attal, Attal2, Karevski} for the whole range of parameters. This requires external power \cite{Barra,Esposito rep. int.}, which means that, just from the Lindblad equation, one cannot safely state that it is an autonomous refrigerator.

The second purpose of this paper is to extend the results of our previous work, on the equilibrium properties of the repeated interaction scheme \cite{StochPRE} when the system dissipates to a single bath, to the case where many baths are coupled to the system. We show under what conditions the thermodynamics can be exactly calculated using only the system degrees of freedom, and show that this corresponds, in the case of Lindblad dynamics, to the condition of \emph{quantum local detailed balance} with respect to a steady-state which might not necessarily be the canonical thermal state. Remarkably, when this condition is fulfiled, the external power vanishes in the steady-state. We show that the smallest possible refrigerator satisfies this condition, thus even though, our derivation indicates that it is non-autonomous in general, the derived thermodynamics in the steady-state is consistent with the results of \cite{linden,linden2}.

To introduce the notation and to make the presentation as self-contained as possible we start with a summary of quantum thermodynamics and previous results. Sections \ref{subsec: thermo open quantum system} and \ref{subsec: Rep. Int.} introduce the thermodynamics of open quantum systems as well as the repeated interaction scheme when there is only one bath. It is shown on section \ref{subsec: concatenation} how this scheme can be though of as a concatenation of maps with or without equilibrium as well as the consequences of these properties. In section  \ref{subsec: Lindblad limit} it is shown how to obtain Lindblad dynamics starting from the repeated interaction scheme. The new contributions start in section \ref{subsec: generalization many baths}, where a generalization of the previous results for many baths is given and the condition of quantum local detailed balance is derived. In section \ref{sec: refrigerator} we summarize the principal aspects of the three-qubit refrigerator model, then we show how to derive its Lindblad equation starting from a repeated interaction scheme and finally we show that it can operate without external power in the steady-state. Conclusions are presented in section \ref{sec: conclusions}.

\section{Thermodynamics in the repeated interaction scheme and the Lindblad limit} \label{sec: seccion 1}

\subsection{Thermodynamics of open quantum systems} \label{subsec: thermo open quantum system}
We begin introducing the notation and concepts of an open quantum system and its thermodynamics. Consider a system $S$ and a heath bath $B$. We will assume that at the initial time ($t=0$) they are decorrelated, forming a product state
\begin{equation}
\rho(0) = \rho_S(0) \otimes \rho_B(0),
\end{equation}
where $\rho_S$ $(\rho_B)$ is the reduced state of the system (bath). In the subsequent time, the system couples to the bath and the total state evolves unitarily according to the Liouville von-Neumann equation $\dot \rho(t) = -i[H(t),\rho(t)]$ ($\hbar = 1$ hereinafter), where the total Hamiltonian is given by
\begin{equation}
H(t) = H_S + H_B + V(t).
\end{equation}
Here, $V(t)$ is the coupling between the system and the bath and it is the only time-depending part of the Hamiltonian. For the repeated interaction scheme the system and bath Hamiltonians are not necessarily time-independent, but in this work we want to emphasize on this case for reasons that will become evident later on when we explore the Lindblad limit.

The total state at time $t$ can be written as
\begin{equation} \label{ecc: evolucion total}
\rho(t) = U(t) \rho_S(0) \otimes \rho_B(0) U(t)^\dag,
\end{equation}
where $U(t) = \mathcal T_+ e^{-i\int_0^t ds H(s)}$ is the unitary time evolution operator and $\mathcal T_+$ denotes the time ordering operator. The states of the system and bath are
$\rho_S(t) = \Tr_B[\rho(t)]$ and $\rho_B(t)=\Tr_S[\rho(t)]$, respectively. If the bath is initially prepared in the Gibbs state with a given inverse temperature $\beta$ ($k_B = 1$ hereinafter), i.e., $\rho_B(0) = e^{-\beta H_B}/Z_B \equiv \omega_\beta(H_B)$, it is possible to study the thermodynamics of this evolution. One starts with the time derivative of the thermodynamic quantities, where the interpretation is clear, and then integrates to obtain the changes during a finite time evolution. The energy change due to an external driving is identified as the power ($\dot W(t)=\Tr[\rho(t)\dot H(t)]$) and the heat flux is chosen as the negative change of the bath's energy ($\dot Q(t) = -\Tr_B[H_B\dot \rho_B(t)]$). With this two choices, as well as the choice to use the von-Neumann entropy ($S(t)=-\Tr_S[\rho_S(t)\ln \rho_S(t)]$), one can write the energy and entropy balance for an open quantum system \cite{Esposito-NJP}:
\begin{equation}
\dot E(t) = \dot W(t) + \dot Q(t), \qquad \frac{d_i}{dt} S = \dot S(t) - \beta \dot Q(t),
\end{equation}
with $E(t)=\Tr[(H_S+V(t))\rho(t)]$. The first corresponds to the first law of thermodynamics and the second law is expressed by the inequality in Eq.(\ref{ecc: entropy production}) below.

In the repeated interaction scheme we will discuss below, there are many identical copies of the bath that interact sequentially, one by one, with the system. We will be interested in the driving arising from turning on and off the interaction with each bath. To carefully take this into account, we write the coupling as
\begin{equation} \label{ecc:definicion V(t)}
V(t)=\Theta(t)\Theta(\tau-t)V,
\end{equation} with a constant coupling operator $V$, a step function $\Theta(t)$ which is equal to $0$ if $t<0$ and $1$ if $t\geq 0$, and $\tau$ the time duration of the coupling with the bath. Integrating the thermodynamic quantities between $t=0$ and $t=\tau$ one obtains
\begin{eqnarray}
&&\Delta Q = -\Tr_B[H_B(\rho_B(\tau)-\omega_\beta(H_B))], \label{ecc: heat}\\
&&\Delta W = \Tr[(H_S+H_B) (\rho(\tau)-\rho(0))], \label{ecc: work}\\
&&\Delta E = \Delta Q + \Delta W = \Tr[H_S(\rho_S(\tau) - \rho_S(0))], \label{ecc: energy change} \\
&&\Delta S = -\Tr_S[\rho_S(\tau) \ln \rho_S(\tau)] + \Tr_S[\rho_S(0) \ln \rho_S(0)], \label{ecc: entropy change}\\
&&\Delta_i S = \Delta S - \beta Q =  D(\rho(\tau) || \rho_S(\tau) \otimes \rho_B(\tau)) + D(\rho_B(\tau)||\omega_\beta(H_B)) \geq 0 \label{ecc: entropy production},
\end{eqnarray}
where $D(a||b) \equiv \Tr[a\ln a]-\Tr[a\ln b] \geq 0$ is the relative entropy. The total final state $\rho(\tau)$ is given by Eq.(\ref{ecc: evolucion total}) with $U(\tau) = e^{-i \tau(H_S+H_B+V)}$.

\subsection{Repeated interaction} \label{subsec: Rep. Int.}
We now consider the case where there are many identical copies of the bath $B$ that interact sequentially with the system. We will consider that each bath has the same Hamiltonian $H_B$ and are initially prepared in the same Gibbs state $\omega_\beta(H_B)$. Each bath will couple to the system for a time lapse $\tau$ with the same coupling strength and, once the coupling is turned off, it will never interact again with the system. We will distinguish between one copy of the bath and another using the super-index $n$ for the state, $\rho_B^n(t)$, indicating that it is the $n$th copy of the bath. Similarly, we will write $H_B^n$ for the Hamiltonian of the $n$th copy of the bath, and $V^n$, the coupling operator in Eq.(\ref{ecc:definicion V(t)}), to indicate that it couples the system with the $n$th copy of the bath. Following references \cite{Attal, Barra} a recursion for the system state is found:
\begin{equation} \label{ecc: system state recursion}
\rho_S(n\tau) = \Tr_B[U^n \rho_S((n-1)\tau) \otimes \omega_\beta^n(H_B^n) U^{n\dag}],
\end{equation}
with $U^n \equiv U^n(\tau) = e^{-i\tau(H_S+H_B^n + V^n)}$. One can see that the repeated interaction induces Markovian dynamics on the system, since it is equivalent to refreshing a single bath to the thermal state periodically with a period $\tau$. Note that the index $n$ in $U^n$, $H_B^n$, $V^n$ and $\rho_B^n$ is only necessary to distinguish between different copies of the bath, i.e., to different Hilbert spaces $\mathcal H_B^n$ in $\mathcal H = \mathcal H_S \otimes \mathcal H_B^1 \otimes \mathcal H_B^2 \otimes \cdots$. Besides from this, the operators have the same form, so we drop the index $n$ on them keeping in mind that at the beginning of each interaction a new bath is used. We rewrite Eq.(\ref{ecc: system state recursion}) as $\rho_S(n\tau) = \Tr_B[U\rho_S((n-1)\tau)\otimes \omega_\beta(H_B)U^\dag]$.

In each iteration there will be a contribution to the thermodynamic quantities. Eqs.(\ref{ecc: heat}--\ref{ecc: entropy production}) take the following form for the $n$th time step:
\begin{eqnarray}
&&\Delta Q^n = -\Tr_B[H_B(\rho_B'-\omega_\beta(H_B))], \label{ecc: heat rep.int.}\\
&&\Delta W^n = \Tr[(H_S+H_B) (\rho(n\tau)-\rho((n-1)\tau))], \label{ecc: work rep.int.} \\
&&\Delta E^n = \Delta Q^n + \Delta W^n = \Tr[H_S(\rho_S(n\tau) - \rho_S((n-1)\tau))], \label{ecc: energy change rep.int.} \\
&&\Delta S^n = -\Tr_S[\rho_S(n\tau) \ln \rho_S(n\tau)] + \Tr_S[\rho_S((n-1)\tau) \ln \rho_S((n-1)\tau)], \label{ecc: entropy change rep.int.}\\
&&\Delta_i S^n = \Delta S^n - \beta Q^n =  D(\rho(n\tau) || \rho_S(n\tau) \otimes \rho_B') + D(\rho_B'||\omega_\beta(H_B)) \geq 0 \label{ecc: entropy production rep.int.},
\end{eqnarray}
where $\rho_B'$ refers to the state of the $n$th copy of the bath at the end of the interaction with the system.

\subsection{Concatenation of CPTP maps} \label{subsec: concatenation}
The repeated interaction can also be studied in terms of concatenation of completely positive trace preserving (CPTP) maps \cite{MHP}, which gives further insight. For a single iteration we define the map $\mathcal E(\cdot)=\Tr_B[U(\cdot)\otimes \omega_\beta(H_B) U^\dag]$. This map preserves the positivity and the trace. In our case, the recursion in Eq.(\ref{ecc: system state recursion}) can be written as $\rho_S(n\tau)= \mathcal E(\rho_S((n-1)\tau))$. This means that the state of the system at time $n\tau$ is a concatenation of $n$ maps $\mathcal E$ applied to the initial state:
\begin{equation}
\rho_S(n\tau) = \underbrace{\mathcal E \mathcal E \cdots \mathcal E }_\text{$n$ times}(\rho_S(0))\equiv \mathcal E^n(\rho_S(0)).
\end{equation}

If we assume that $\mathcal E$ has a unique invariant state, $\pi=\mathcal E(\pi)$, attractive due to the contractive character of the relative entropy under the action of the map~\cite{Breuer},
% and $\lim_{N\to\infty} \mathcal E^N(\rho)=\pi$ $\forall \rho$, 
 we can distinguish two kind of maps, maps with or without equilibrium. If the action of $\mathcal E$ over $\pi$ gives $\Delta_i S > 0$ then we say it is a \textit{map without equilibrium} or \textit{map with a NESS} (non-equilibrium steady state). Conversely, if the action of the map over $\pi$ gives $\Delta_i S=0$, then we call it a \textit{map with equilibrium}~\cite{StochPRE}. The unitary time evolution operator $U$ of a map with equilibrium must fulfill 
\begin{equation} \label{ecc: map equilibrium condition}
[U,H_0+H_B]=0
\end{equation}
and the equilibrium state must be of the form $\pi=e^{-\beta H_0}/Z_0$, where $Z_0$ is the partition function and $H_0$ is an operator on the system's Hilbert space, which could be the system Hamiltonian $H_S$ but, most interestingly, it can be something else (in the particular case of the three-qubit absorption refrigerator we study later, $H_0$ is the free Hamiltonian of the three qubits). In the discussion of the Lindblad limit below, we will focus on the cases that $[H_S,H_0]=0$.

Among the properties of maps with equilibrium, the most important one is that the thermodynamic quantities Eqs.(\ref{ecc: heat rep.int.}-\ref{ecc: entropy production rep.int.}) can be written in terms of system operators only. Heat, work and entropy production take the form
\begin{eqnarray}
&&\Delta Q^n = \Tr_S[H_0(\rho_S(n\tau)-\rho_S((n-1)\tau))], \label{ecc: heat local} \\
&&\Delta W^n = \Tr_S[(H_S-H_0)(\rho_S(n\tau)-\rho_S((n-1)\tau)], \label{ecc: work local}\\
&&\Delta_i S^n = D(\rho_S((n-1)\tau)||\pi) - D(\rho_S(n\tau)||\pi) \geq 0. \label{ecc: entropy production local}
\end{eqnarray}
We say in this case that the \textit{thermodynamics is local}, in the sense that it depends on local (system) quantities. On the contrary, if the quantities depend on the system and bath, we say the \textit{thermodynamics is global}.

\subsection{Lindblad limit} \label{subsec: Lindblad limit}

It is possible in the repeated interaction to obtain in the continuous limit a Lindblad equation for the dynamics of the system state. As discussed in \cite{Attal, Attal2, Karevski}, one scales the system-bath coupling as $V=v/\sqrt{\tau}$, then takes the limits $\tau \to 0$ and the number of the copies of the bath (or, equivalently, the number of CPTP maps) to infinity, in such a way that $t=n\tau$ is finite. One obtains
\begin{equation} \label{ecc: Lindblad equation}
\dot \rho_S(t) \equiv \lim\limits_{\substack{\tau \to 0 \\ n \to \infty}} \frac{\rho_S(n\tau) - \rho_S((n-1)\tau)}{\tau} =  -i[H_S,\rho_S(t)] + \mathcal D(\rho_S(t)),
\end{equation}
where the dissipator $\mathcal D$ is
\begin{equation}
\mathcal D(\rho_S(t)) = \sum_k \gamma_k\left( L_k \rho_S(t) L_k^\dag - \frac{1}{2}\{ L_k^\dag L_k, \rho_S(t)\} \right).
\end{equation}
To obtain this Lindblad equation, the standard condition $\Tr_B[V\omega_\beta(H_B)]=0$ must be fulfilled. Here, the label $k$ corresponds to two labels, $k=\{ij\}$. The rate $\gamma_{k}$ is given by $\bra{i}\omega_\beta(H_B)\ket{i}$ and the Lindblad operator $L_k$ is $\bra{j}v\ket{i}$, where $\ket{i}$ and $\ket{j}$ are eigenstates of $H_B$. One can verify that the invariant state of the Lindblad equation is the same invariant state $\pi$ of the concatenated map $\mathcal E$. 

In this limit, the thermodynamic quantities of Eqs.(\ref{ecc: heat rep.int.}-\ref{ecc: entropy production rep.int.}) take the continuous form \cite{Barra}
\begin{eqnarray}
&&\dot Q(t) \equiv \lim \frac{\Delta Q^n}{\tau} = -D(H_B), \label{ecc: heat Lindblad}\\
&&\dot W(t) \equiv \lim \frac{\Delta W^n}{\tau} = D(H_S + H_B), \label{ecc: work Lindblad}\\
&&\dot E(t) \equiv \lim \frac{\Delta E^n}{\tau} = \Tr_S[H_S \mathcal D(\rho_S(t))], \label{ecc: energy change Lindblad} \\
&&\dot S(t) \equiv \lim \frac{\Delta S^n}{\tau} = -\Tr_S[\dot \rho_S(t) \ln \rho_S(t)], \label{ecc: entropy change Lindblad}\\
&&\frac{d_i}{dt} S(t) \equiv \lim \frac{\Delta_i S^n}{\tau} = \dot S(t) - \beta \dot Q(t),\label{ecc: entropy production Lindblad}
\end{eqnarray}
where the limit refers to $\tau \to 0$ and $n\to \infty$. Note that heat flux, power and entropy production rate are determined by $D$, which is given by
\begin{equation}
D(A) = \Tr\left[\left(vAv - \frac{1}{2}\{v^2,A\}\right) \rho_S(t) \otimes \omega_\beta(H_B)\right].
\end{equation}
This expression depends on the bath and the system-bath coupling. One thus needs more information than what is given by the Lindblad equation to compute the thermodynamic quantities. This is indeed the case when $\mathcal E$ is a map  with NESS. The notion of global thermodynamics for maps without equilibrium is also present in the Lindblad equations with NESS.

If, instead, a concatenation of a map $\mathcal E$ with equilibrium is used, then the resulting Lindblad equation has
an equilibrium steady state given by the equilibrium of the map, $\pi=e^{-\beta H_0}/Z_0$. To derive the heat flux, power and entropy production rate in this case, one can either take the continuous limit of Eqs.(\ref{ecc: heat local}-\ref{ecc: entropy production local}) or use the equilibrium condition of a map with equilibrium on Eqs.(\ref{ecc: heat Lindblad},\ref{ecc: work Lindblad},\ref{ecc: entropy production Lindblad}). In both ways one arrives to the same result. For later discussion we will use the equilibrium condition, Eq.(\ref{ecc: map equilibrium condition}), which, together with the assumption $[H_S,H_0]=0$, reduces to
\begin{equation}
[V,H_0+H_B]=0.
\end{equation}
With this commutation relation one can readily check that $D(H_B) = -D(H_0)$. Thus, $\dot Q(t) = D(H_0)$ and $\dot W(t) = D(H_S-H_0)$. Noting that for any system operator $O_S$ one has $D(O_S)=\Tr_S[O_S\mathcal D(\rho_S(t))]$, we can write the thermodynamic quantities as
\begin{eqnarray}
&&\dot Q(t) = \Tr_S[H_0 \mathcal D(\rho_S(t))], \\
&&\dot W(t) = \Tr_S[(H_S - H_0) \mathcal D(\rho_S(t))].
\end{eqnarray}
In this case, all the thermodynamic quantities can be written in terms of information contained in the Lindblad equation. We can say then that the notion of local thermodynamics is carried on from the maps to the Lindblad equation.

A remarkable property of the Lindblad equation with the equilibrium state $\pi=e^{-\beta H_0}/Z_0$ is that it satisfies the quantum detailed balance condition with respect to $\pi$~\cite{Alicki76,StochPRE}:
\begin{equation}
\mathcal L_a(\pi)=0, \qquad \mathcal L_s(A\pi) = \mathcal L_s^*(A)\pi,
\end{equation} 
for every bounded operator $A$. Here $\mathcal L_a$ is the anti-hermitian part of the Lindbladian, i.e., the part of Eq.(\ref{ecc: Lindblad equation}) corresponding to the unitary dynamics; $\mathcal L_s$ is the hermitian part, i.e., the dissipator $\mathcal D$, and $\mathcal L_s^*$ is its dual.

It is important to note that this microscopic derivation of the Lindblad dynamics gives an expression, Eq.(\ref{ecc: Lindblad equation}), where work (power) is `hidden', in the sense that neither the system Hamiltonian nor the dissipator depend on time. 
For this reason, one should be careful with thermodynamic statements for a process described by a Lindblad dynamics without explicitly mentioning its underlying microscopic derivation.

\subsection{Generalization to many baths} \label{subsec: generalization many baths}

Some of the previous results can be generalised to the case of a system $S$ interacting at the same time with different baths $B_1$, $B_2$, $\dots$, $B_N$ with equal or different temperatures, and through equal or different couplings. We will use the index $r\in\{1,\dots,N\}$ to label the bath and its coupling. Replacing $\rho_B(t)$ and $H_B$  by $\bigotimes_r \rho_{B_r}(t)$ and $\sum_r H_{B_r}$, respectively, in Eqs.(\ref{ecc: heat rep.int.}) and (\ref{ecc: work rep.int.}), one obtains for the repeated interaction that the total heat and work for the $n$th iteration are given by
\begin{eqnarray}
&&\Delta Q^n = \sum\limits_r \Delta Q^n_r = \sum\limits_r -\Tr_{B_r}[H_{B_r}(\rho_{B_r}'-\omega_{\beta_r}(H_{B_r}))], \label{ecc: heat rep.int. many baths}\\
&&\Delta W^n = \sum\limits_r \Delta W^n_r = \sum\limits_r \Tr[(H_S+H_{B_r})(\rho(n\tau) - \rho((n-1)\tau))], \label{ecc: work rep.int. many baths}
\end{eqnarray}
There is now a heat and work contribution for each bath and the respective coupling and $Q^n_r$ ($W^n_r$) is the heat (work) contribution of the $n$th copy of the bath $r$. 

In the Lindblad limit, the equation obtained is
\begin{equation} \label{ecc: Lindblad many baths}
\dot \rho_S(t) = -i[H_S,\rho_S(t)] + \sum_r \mathcal D_r(\rho_S(t)),
\end{equation}
where
\begin{equation}
\mathcal D_r(\rho_S(t)) = \sum\limits_{k_r} \gamma_{k_r} \left(L_{k_r} \rho_S(t) L_{k_r}^\dag - \frac{1}{2}\{ L_{k_r}^\dag L_{k_r},\rho_S(t) \} \right).
\end{equation}
Here, $k_r=\{i_r,j_r\}$, $\gamma_{k_r} = \bra{i_r}\omega_{\beta_r}(H_{B_r})\ket{i_r}$ and $L_{k_r} = \bra{j_r}v_{r}\ket{i_r}$, just as before. The heat flux and power, Eqs.(\ref{ecc: heat Lindblad}) and (\ref{ecc: work Lindblad}), take the form
\begin{eqnarray}
&&\dot Q(t) = \sum\limits_r \dot Q_r(t) = \sum\limits_r -D_r(H_{B_r}), \label{ecc: heat Lindblad many baths} \\
&&\dot W(t) = \sum\limits_r \dot W_r(t) = \sum\limits_r D_r(H_S + H_{B_r}), \label{ecc: work Lindblad many baths}
\end{eqnarray}
where
\begin{equation}
D_r(A) = \Tr\left[ \left( v_r A v_r - \frac{1}{2}\{v_r^2,A\} \right) \rho_S(t) \otimes \omega_{\beta_r}(H_{B_r}) \right].
\end{equation}
Note that $\Tr[v_r \omega_{\beta_r}(H_{B_r})]=0$ has been used in Eqs.(\ref{ecc: work rep.int. many baths}), (\ref{ecc: Lindblad many baths}) and (\ref{ecc: work Lindblad many baths}) to separate each bath contribution.

One may wonder whether the thermodynamics could be local when there are many baths, with different temperatures, present. To answer this question, consider the following. If $U=e^{-i\tau(H_S+\sum_{r=1}^N (H_{B_r} +V_r))}$ is the unitary time evolution operator governing the dynamics of the system and the $N$ different baths, we define as $U_r = e^{-i \tau(H_S + H_{B_r} + V_r)}$ the evolution operator that each of these baths plus the system would have if they were alone, interacting one with the other. From the set of baths $\{1,2,\dots,N\}$, we define as $\mathcal C$ the subset of baths which would have associated a map with a given equilibrium if were coupled alone to the system. This is, baths $r\in C$ fulfil the following equilibrium condition:
\begin{equation}
[U_r, H_0+H_{B_r}] =0 \qquad \text{if $r\in \mathcal C$},\label{ecc:cond-eq}
\end{equation}
for an arbitrary system operator $H_0$. The rest of the baths, that are associated to maps with NESS, fulfil
\begin{equation}
[U_r,O_S + H_{B_r}] \neq 0 \qquad \text{if $r \not \in \mathcal C$},
\end{equation}
for any system operator $O_S$.

With these considerations, together with $[H_S,H_0]=0$, we have that $[V_r,H_0+H_{B_r}]=0$ if $r\in \mathcal C$. 
%Thus, heat and work in the repeated interaction, Eqs.(\ref{ecc: heat rep.int. many baths}) and (\ref{ecc: work rep.int. many baths}), separate into terms depending solely on the system and terms depending on the system and baths:
%\begin{eqnarray}
%&&\Delta Q^n = \sum\limits_{r\in \mathcal C} \Tr_S[H_0(\rho_S(n\tau) - \rho_S((n-1)\tau))]  + \sum\limits_{r\not \in \mathcal C} \Tr_{B_r}[H_{B_r}(\rho_{B_r}'-\omega_{\beta_r}(H_{B_r}))], \\
%&&\Delta W^n = \sum\limits_{r\in \mathcal C} \Tr_s[(H_S-H_0)(\rho_S(n\tau) - \rho_S((n-1)\tau))] +\sum\limits_{r \not \in \mathcal C} \Tr[(H_S+H_{B_r})(\rho(n\tau) - \rho((n-1)\tau))].
%\end{eqnarray}
%Note the terms appearing in the first summation of both equations do not depend on $r$, thus $\sum_{r\in \mathcal C} = |\mathcal C|$.
The contributions of a bath $r\in C$ to heat and work in the Lindblad limit can be written in terms of system operators. Heat flux and power, Eqs.(\ref{ecc: heat Lindblad many baths}) and (\ref{ecc: work Lindblad many baths}), become
\begin{eqnarray}
&&\dot Q(t) = \sum\limits_{r \in \mathcal C} D_r(H_0) + \sum\limits_{r \not \in \mathcal C} -D_r(H_{B_r}) = \sum\limits_{r \in \mathcal C} \Tr_S[H_0 \mathcal D_r(\rho_S(t))] + \sum\limits_{r \not \in \mathcal C} -D_r(H_{B_r}), \\
&& \dot W(t)= \sum\limits_{r \in \mathcal C} D_r(H_S-H_0) + \sum\limits_{r\not \in \mathcal C} D_r(H_S+H_{B_r}) = \sum\limits_{r \in \mathcal C} \Tr_S[(H_S-H_0)\mathcal D_r(\rho_S(t))] + \sum\limits_{r\not \in \mathcal C} D_r(H_S+H_{B_r}).
\end{eqnarray}
In the particular case that every bath can be associated to an equilibrium map if connected alone to the system, this is, $\mathcal C=\{1,\dots,N\}$, all the thermodynamic quantities can be written in local form. Heat flux and power are~\footnote{Note that one could have a solution $H_0^r$ of Eq.(\ref{ecc:cond-eq}) but we restrict ourselves to the case $H_0^r=H_0\,\forall r$ because in that case in the steady state the power vanishes, see below.}
\begin{eqnarray}
&&\dot Q(t) = \sum_r \dot Q_r(t) = \sum_r \Tr_S[H_0 \mathcal D_r(\rho_S(t))], \\
&&\dot W(t) =\sum_r \dot W_r(t) = \sum_r \Tr_S[(H_S-H_0) \mathcal D_r(\rho_S(t))].
\end{eqnarray}
In the NESS (there are different temperatures so, in general, the system cannot equilibrate), the total heat flux and power vanish. It can be readily verified noting that $\sum_r \mathcal D_r(\rho_S^\text{NESS}) = i[H_S,\rho_S^\text{NESS}]$ (see Eq.(\ref{ecc: Lindblad many baths})). Thus,
\begin{eqnarray}
&& \dot Q^\text{NESS} = i \Tr_S[[H_0,H_S]\rho_S^\text{NESS}]=0, \\
&&\dot W^\text{NESS} = i\Tr_S[[H_S-H_0,H_S]\rho_S^\text{NESS}=0. \label{ecc: work ness local detailed balance}
\end{eqnarray}
Both quantities vanish due to $[H_S,H_0]=0$. This is the central result of this section, it is a generalization of \emph{quantum local detailed balance} because the `local equilibriums' are given by $\omega_{\beta_i}(H_0)$ and not by $\omega_{\beta_i}(H_S)$ \cite{Spohn LDB}. Note that if all temperatures are equal ($\beta_i=\beta$), the system would thermalize to $\omega_{\beta}(H_0)$. This fact is normally considered as an inconsistency of the local approach in the weak-coupling regime \cite{Levy-Kosloff, Correa2017, Marti2017, Plenio2017}, however, in the context of the repeated interaction the Markovian master equation is exact, thus, this non-standard thermalization is the expected result. We offer an intuitive argument for our example at the end of Section \ref{sec: refrigerator}.

In the following section we will summarize the principal aspects of the three qubit refrigerator model and then we will show how to derive its Lindblad equation from a particular repeated interaction model. Then we will use the result of Eq.(\ref{ecc: work ness local detailed balance}) to argue that the refrigerator does not require external power in the stationary state (where it operates), and thus, can reach Carnot's coefficient of performance.

\section{Three-qubit absorption refrigerator} \label{sec: refrigerator}

\subsection{The model}

The smallest possible absorption refrigerator was studied in \cite{linden} where three models were proposed. Among the three of them, the one that has had more attention is the three-qubit refrigerator. This model consists of a system composed of three qubits, each one connected to a different bath. The idea is that in the stationary state of the system, one of the qubits reaches an effective stationary temperature lower than that of the bath to which is coupled. In this way, the qubit continuously extracts energy from its bath, cooling it.

To mathematically describe this refrigerator, we introduce the notation. We will use the sub-index $i$ to label one of the three qubits,  $i=\{1,2,3\}$, and its corresponding bath. The first qubit is coupled to the \textit{cold} bath ($\beta_1^{-1} = T_C$); the second one, to the \textit{room} bath ($\beta_2^{-1} = T_R$); and the third one, to the \textit{hot} bath ($\beta_3^{-1} = T_H$). Temperatures are chosen in the order $T_C<T_R<T_H$. It is the cold bath the one which is cooled in the stationary state.

The Hamiltonian of the system is given by $H_S=H_0+H_\text{int}$ where the first term is the free Hamiltonian of the qubits,
\begin{equation}
H_0 = \sum\limits_{i=1}^3 H_i = \sum\limits_{i=1}^3 E_i \ket{1}_i\bra{1}_i, \label{ecc: free Hamiltonian refri}
\end{equation}
and the second term is the interaction between the qubits,
\begin{equation}
H_\text{int} = g(\ket{010}\bra{101} + \ket{101}\bra{010}). \label{ecc: int Hamiltonian refri}
\end{equation}
Here $\ket{1}_i$ and $E_i$ are the excited state and energy of the $i$-qubit. The energies are chosen such that $E_2=E_1+E_3$; in this way, $H_\text{int}$ couples degenerated states.

An intuitive reset model is used to describe the dissipation in the baths' presence. With probability $p_i$ per time $\delta t$, each qubit may reset to a thermal state with the temperature of its bath. This is, of course, induced by the latter. Including the unitary part of the dynamics, the limit $\delta t \to 0$ leads to the master equation
\begin{equation} \label{ecc: Master equation refrigerator}
\dot \rho_S(t) = -i[H_S,\rho_S(t)] + \sum\limits_{i=1}^3 p_i\left( \omega_{\beta_r}(H_i)\Tr_i[\rho_S(t)] - \rho_S(t)\right),
\end{equation}
where $\omega_{\beta_i}(H_i) = e^{-\beta_i H_i}/\Tr[e^{-\beta_r H_i}]$. This master equation can be rewritten in Lindblad form as
\begin{equation} \label{ecc: Lindblad equation refrigerator}
\dot \rho_S(t) = -i[H_S,\rho_S(t)] + \sum\limits_{i=1}^3 \sum\limits_{k=1}^4 \gamma_i^k \left(L_i^k \rho_S(t) L_i^{k\dag} - \frac{1}{2}\{L_i^{k\dag}L_i^k,\rho_S(t)\} \right),
\end{equation}
where $\gamma_i^{1,2} = p_i\bar r_i$, $\gamma_i^{3,4}=p_i r_i$, $L_i^1 = \sigma_i^+$, $L_i^2=\sigma_i^+\sigma_i-$, $L_i^3 = \sigma_i^-$ and $L_i^4=\sigma_i^- \sigma_i^+$. Here we have introduced the quantities $r_i = \bra{0}_i\omega_{\beta_i}(H_i)\ket{0}_i$ and $\bar{r}_i = \bra{1}_i \omega_{\beta_i}(H_i)\ket{1}_i$, while $\sigma_i^\pm = (\sigma_i^x \pm i\sigma_i^y)/2$. It is argued \cite{linden,linden2} that the inclusion of $H_\text{int}$ would require modifications to the dissipator, but in the limits $g\to0$ and $p_i\to 0$, such that $g/p_i$ remains constant, the corrections are of order $gp_i$.

The stationary state solution can be found analytically \footnote{The analytical steady state solution in \cite{linden2}, and in the second version of the arXiv, has two typos. The correct expression can be found in the first version, arXiv:1009.0865v1} and is exact for any value of $g$ and $p_i$. Tracing over two qubits, the stationary state of qubit $i$ is given by:
\begin{equation} \label{ecc: rho_i ness refrigerator}
\rho_i^\text{NESS} = \omega_{\beta_i}(H_i) + \frac{q \gamma}{p_i} (-1)^{i}\sigma_i^z,
\end{equation}
where $q=p_1+p_2+p_3$ and $\gamma$ is a complicated expression which is not relevant for our purposes, except for the fact that it determines the cooling window:
\begin{equation} \label{ecc: cooling window}
0<\frac{E_1}{E_3} < \frac{(T_H-T_R)T_C}{(T_R-T_C)T_H}.
\end{equation}

For this autonomous refrigerator, since there is no external work, the coefficient of performance (COP) can be computed in terms of the heat fluxes, which equal the energy changes of each qubit. These are given by
\begin{equation} \label{ecc: heat fluxes linden et al}
\dot Q_C = q \gamma E_1, \quad \dot Q_R = -q\gamma E_2, \quad \dot Q_H = q \gamma E_3.
\end{equation}
Thus, the COP is
\begin{equation}
\eta \equiv \frac{\dot Q_C}{\dot Q_H} = \frac{E_1}{E_3},
\end{equation}
which, according to Eq.(\ref{ecc: cooling window}), can reach the upper bound $(T_H-T_R)T_C/(T_R-T_C)T_H$ which is exactly Carnot's COP \cite{linden2}. This limit is reached asymptotically for vanishing \textit{cooling power}.

It was noticed in~\cite{linden} that to justify the model, the limit $g\to 0$ is required. However, a later work \cite{Correa} argued that the local dissipation in Eq.(\ref{ecc: Master equation refrigerator}) (or (\ref{ecc: Lindblad equation refrigerator})) is not physically possible. In particular, it was shown that a  microscopically derived Lindblad equation for this refrigerator in the Born-Markov-Secular approximation, has de-localized dissipation that cannot be neglected when $g$ and $p_i$ are non-zero, however small they may be. With their `consistent' Lindblad equation they obtained a lower bound for the coefficient of performance. However, recent progress has been made on justifying the local approach without using the secular approximation~\cite{Correa2017, Marti2017, Plenio2017}.
%  and we agree with them that it can correctly describe the dynamics of the real system {\color{blue}in a particular regime}. Nevertheless, we would like to argue that the validity of the local approach is not enough to make thermodynamic statements if no microscopic derivation is specified.

In the following we will show how Eq.(\ref{ecc: Lindblad equation refrigerator}) can be microscopically derived within the repeated interaction scheme without any assumptions on $g$ or $p_i$. It can be anticipated that in this case the refrigerator will be non-autonomous because of the work needed to switch on and off the couplings. However, as mentioned in the previous section, we will show that Eq.(\ref{ecc: Lindblad equation refrigerator}) has quantum local detailed balance (because the internal interaction is energy preserving $[H_0,H_{\rm int}]=[H_0,H_S]=0$), thus the total power vanishes in the stationary state.

\subsection{Repeated interaction derivation}

\begin{figure}[H] 
\centering
  \includegraphics[width=0.4\textwidth]{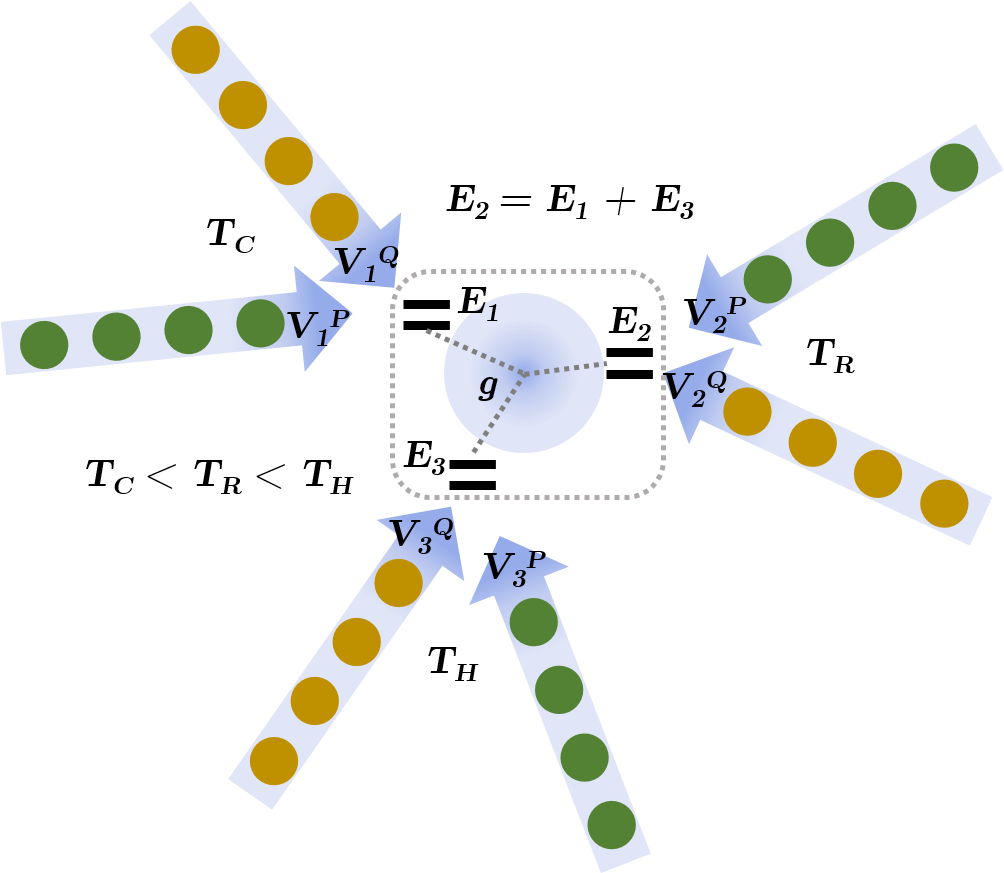}
    \caption{(Color online) Schematic diagram of the three-qubit refrigerator. Two streams of identical copies of a bath alternately interact with each qubit. Gold circles represent the copies which couple to the system through a potential $V_i^Q$ in odd time steps, while green circles interact through $V_i^P$ in even time steps.}
    \label{fig:esquema}
\end{figure}

To construct proper baths and couplings, so that Eq.(\ref{ecc: Lindblad equation refrigerator}) is obtained (for arbitrary $g$) from a repeated interaction, we will consider a slightly more intricate scheme that the one in Section \ref{subsec: Rep. Int.}. We propose as an  \textit{anzat} that each qubit couples alternately to two streams of copies of its corresponding bath, each stream consisting on thermal qubits whose Hamiltonian are specified later. These couplings are given by
\begin{equation}
V_i^Q = J_i\left(\sigma_{B_i}^x\sigma_i^x + \sigma_{B_i}^y\sigma_i^y + Q_i(\sigma_{B_i}^z - M_i)\sigma_i^z \right), \qquad V_i^P = J_i \left(\sigma_{B_i}^x\sigma_i^x + \sigma_{B_i}^y\sigma_i^y + P_i(\sigma_{B_i}^z - M_i)\sigma_i^z \right).
\end{equation}
Here $\sigma_i^{x,y,z}$ are the Pauli spin 1/2 operators of the system qubit $i$, and $\sigma_{B_i}^{x,y,z}$ the ones of the bath associated to that qubit. The coefficients $Q_i$ and $P_i$ are unknown for the moment and are part of the \textit{anzat}. $J_r$ is a coupling energy, which will be later fixed, and $M_i=\Tr_r[\sigma_{B_i}^z \omega_{\beta_i}(H_{B_i})]$. In Fig. \ref{fig:esquema} we show a schematic drawing of the system and the six streams of bath copies.

In even time steps the system-baths coupling is $V^P \equiv \sum_i V_i^P$ and in odd time steps it is $V^Q \equiv \sum_i V_i^Q$. The state of the total system after the first and second time steps reads
\begin{equation}
\rho(\tau) = U^{1,Q} (\rho_S(0) \otimes \rho_B^1)(U^{1,Q})^\dag \otimes \rho_B^2 \otimes \rho_B^3 \otimes \cdots
\end{equation}
\begin{equation}
\rho(2\tau) = \mathcal U^{2,P} (U^{1,Q} (\rho_S(0) \otimes \rho_B^1)(U^{1,Q})^\dag \otimes \rho_B^2) (\mathcal U^{2,P})^\dag \otimes \rho_B^3 \otimes \cdots,
\end{equation}
where $U^{1,Q}=e^{-i\tau(H_S + H_{B}^1 + V^{1,Q})}$ and $\mathcal U^{2,P} = e^{-i\tau(H_S + H_B^2 + V^{2,P})}e^{-i\tau H_B^1}$. Here, $H_B^n = \sum_{i=1}^3 H_{B_i}^n$ is the Hamiltonian of the $n$th copies of the three baths; $V^{n,Q}=\sum_{i=1}^3 V_i^{n,Q}$ is the total coupling between the system and the $n$th copies of the three baths; and $V^{2n,P} = \sum_{i=1}^3 V_i^{2n,P}$ is the total coupling between the system and the $2n$th copies of the three baths. The $n$th copies of the baths are initially prepared in the thermal states $\rho_B^n = \omega_{\beta_1}(H_{B_1}^n) \otimes \omega_{\beta_2}(H_{B_2}^n) \otimes \omega_{\beta_3}(H_{B_3}^n)$, as in Section \ref{sec: seccion 1}.

If one calculates the state of the total system for further time steps and then trace out the baths, the recursion for the system state that is obtained is
\begin{equation}
\rho_S((2n-1)\tau) = \Tr_{(2n-1)} \left[U^{2n-1,Q}(\rho_S((2n-2)\tau) \otimes \rho_B^{2n-1}) (U^{2n-1,Q})^\dag \right]
\end{equation}
for odd time steps and
\begin{equation}
\rho_S(2n\tau) = \Tr_{(2n)} \left[U^{2n,P}(\rho_S((2n-1)\tau) \otimes \rho_B^{2n}) (U^{2n,P})^\dag \right]
\end{equation}
for even time steps. Here $\Tr_n$ indicates that is the trace over the $n$th copies of the baths.

Scaling the couplings as $V^R=v^R/\sqrt{\tau}$, for $R=\{Q,P\}$, (through the coupling energies $J_i\equiv\sqrt{\lambda_i/\tau}$) we take the limits $\tau\to 0$ and $n\to\infty$, as in Section \ref{subsec: Lindblad limit}, but now fixing $2n\tau = t$, and obtain 
\begin{equation}
\dot \rho_S(t) \equiv \lim\limits_{\substack{\tau \to 0 \\ n \to \infty}} \frac{\rho_S(2n\tau) - \rho_S((2n-2)\tau)}{2\tau} = -i[H_S,\rho_S(t)] + \frac{1}{2}\mathcal D^Q(\rho_S(t)) + \frac{1}{2}\mathcal D^P(\rho_S(t)),
\end{equation}
with 
\begin{equation}
\mathcal D^R(\rho_S(t)) = \Tr_B\left[v^R \rho_S(t) \otimes \rho_B v^R  - \frac{1}{2}\{(v^R)^2,\rho_S(t) \otimes  \rho_B \} \right].
\end{equation}
Here $\rho_B = \omega_{\beta_1}(H_{B_1})\otimes \omega_{\beta_2}(H_{B_2}) \omega_{\beta_3}(H_{B_3})$.

We will now choose the baths to be qubits described by the Hamiltonians $H_{B_i} \equiv E_i \ket{1}_{B_i}\bra{1}_{B_i}$, i.e., they will be equal to their corresponding system qubits. Note that $\Tr_{B_i}[v_i^R \omega_{\beta_i}(H_i)]=0$, thus, splitting the couplings, $v^R=\sum_{i=1}^3 v_i^R$, we can separate the dissipators $Q$ and $P$ into local dissipators:
\begin{equation} \label{ecc: lindblad equation refrigerator from rep. int.}
\dot \rho_S(t) = -i[H_S,\rho_S(t)] + \frac{1}{2}\sum\limits_{i=1}^3 \mathcal D_i^Q(\rho_S(t)) + \frac{1}{2}\sum\limits_{i=1}^3 \mathcal D_i^P(\rho_S(t)),
\end{equation}
with
\begin{equation}
\mathcal D_i^R(\rho_S(t)) = \Tr_{B_i}\left[v_i^R \rho_S(t) \otimes \omega_{\beta_i}(H_{B_i}) v_i^R - \frac{1}{2}\{(v_i^R)^2, \rho_S(t) \otimes \omega_{\beta_i}(H_{B_i}) \} \right].
\end{equation}
To obtain Eq.(\ref{ecc: Lindblad equation refrigerator}) from Eq.(\ref{ecc: lindblad equation refrigerator from rep. int.}), we just need to choose the coefficients of our \textit{anzat} wisely. If $Q_i \equiv (1-M_i)^{-1/2}$, $P_i\equiv(1+M_i)^{-1/2}$ and $\lambda_i \equiv p_i/4$, we recover exactly the Lindblad equation noting that $M_i=\Tr_{B_i}[\sigma_{B_i}\omega_{\beta_i}(H_{B_i})]=1-2r_i$.

In the same manner that we obtained the heat flux and power expressions for the Lindblad limit in Section \ref{sec: seccion 1}, we now obtain, reminding that there are two contributions ($Q$ and $P$) for each bath, the following expressions:
\begin{eqnarray}
&&\dot Q_i \equiv \frac{1}{2}\dot Q_i^Q + \frac{1}{2}\dot Q_i^P = -\frac{1}{2}D_i^Q(H_{B_i}) - \frac{1}{2}D_i^P(H_{B_i}), \\
&&\dot W_i \equiv \frac{1}{2} \dot W_i^Q + \frac{1}{2}\dot W_i^P = \frac{1}{2}D_i^Q(H_S+H_{B_i}) + \frac{1}{2} D_i^P(H_S+H_{B_i}),
\end{eqnarray}
where $D_i^R(A) = \Tr[(v_i^R A v_i^R - \frac{1}{2}\{(v_i^R)^2,A\})\rho_S(t)\otimes \omega_{\beta_i}(H_{B_i})]$.

\begin{figure}[H] 
\centering
  \includegraphics[width=0.4\textwidth]{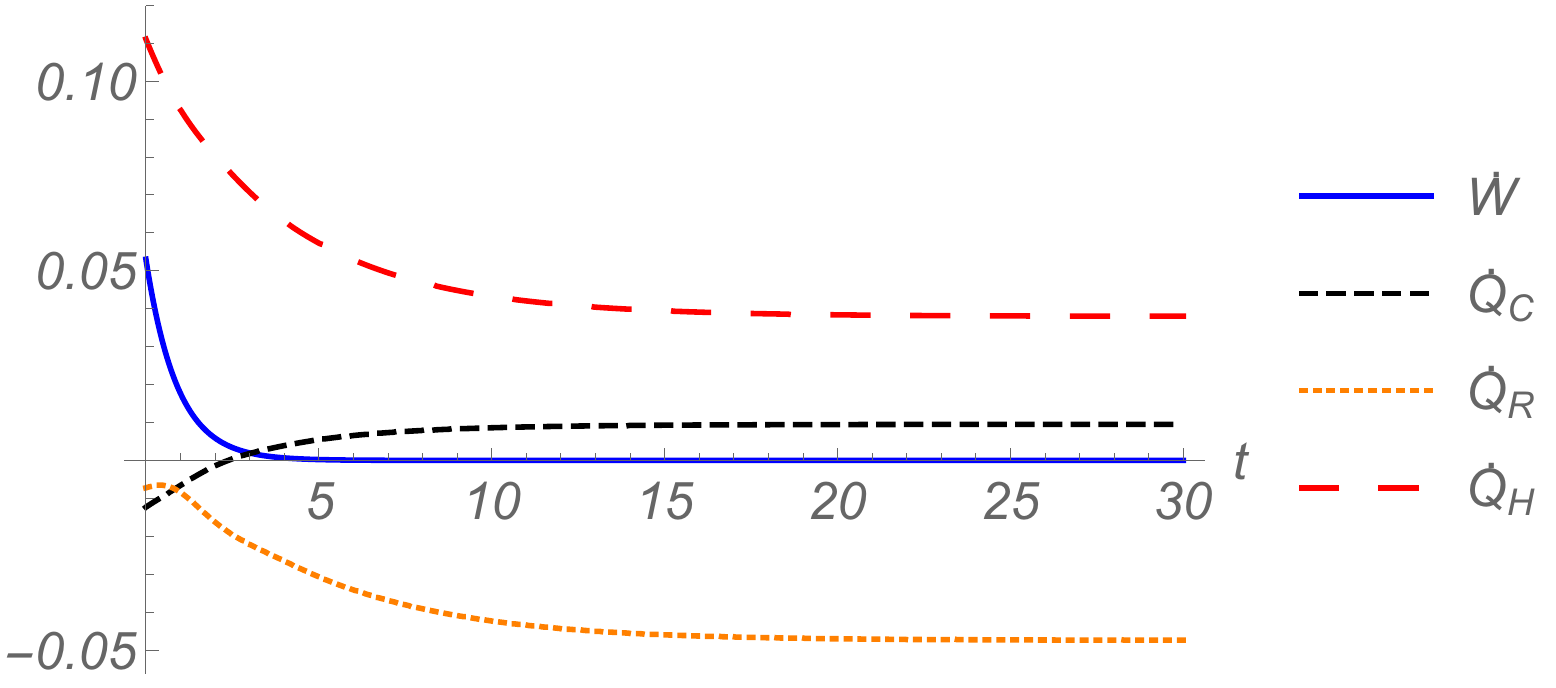}
    \caption{(Color online) .}
    \label{fig:grafico}
    \caption{(Color online) The power (blue) and heat fluxes of the cold (black dashed), room (orange tiny-dashed) and hot (red large-dashed) baths  as a function of time. The values of the parameters are: $E_1 = 1$, $E_2=5$, $E_3=4$, $g=1$, $p_1=0.5$, $p_2=0.4$, $p_3=0.2$, $T_1=2$, $T_2=2.5$ and $T_3=5$, all in units of [time]$^{-1}$.}
\end{figure}

Before evaluating these thermodynamic quantities, note that $[H_S,H_0]=0$ is fulfilled for this system. This is due to the condition $E_2=E_1+E_3$, which, in other words, makes the interaction between the system qubits energy preserving ($[H_S,H_\text{int}]=0$). Additionally, $[v_i^R,H_0+H_{B_i}]=0$ for $i=1,2,3$. Thus, this system has quantum local detailed balance, our main result at the end of Section \ref{sec: seccion 1}. This means that, at the stationary state, where this system was initially conceived to work as a refrigerator, it is indeed an absorption (autonomous) refrigerator because the total power vanishes. Now one might ask whether this microscopic derivation of the Lindblad equation somehow modifies the heat fluxes in the stationary state, and with this, modifies the maximum coefficient of performance it can reach. If one calculates the heat fluxes explicitly, which, due to the generalized local detailed balance, now take the form
\begin{equation}
\dot Q_i = \frac{1}{2}\Tr_S[H_0\mathcal D_i^Q(\rho_S^\text{NESS})] + \frac{1}{2} \Tr_S[H_0 \mathcal D_i^P(\rho_S^\text{NESS})],
\end{equation}
one obtains $\dot Q_i =(p_i/2) E_i\Tr[\sigma_i^z (\omega_{\beta_i}(H_i) - \rho_i^\text{NESS})]$. Using Eq.(\ref{ecc: rho_i ness refrigerator}), this is $\dot Q_i = (-1)^{i+1}q\gamma E_i$, i.e., they are the same than in Eq.(\ref{ecc: heat fluxes linden et al}). Heat fluxes and power are shown if Fig.\ref{fig:grafico} for a non-negligible internal coupling, $g\sim E_i$. In the stationary regime the system works as an absorption refrigerator, extracting heat from the cold bath ($\dot Q_C >0$).

We can conclude that there is a consistent microscopic derivation for the Lindblad equation with local dissipation of the three-qubit refrigerator. It is not autonomous in general because of the nature of the repeated interaction, but in the stationary state it requires no external power. This means that this refrigerator can indeed reach Carnot's coefficient of performance in the limit of vanishing cooling power.

As we commented previously, if all temperatures are equal $\beta_i=\beta$ (single bath), the system would thermalize to $\omega_\beta(H_0)$. This is not what is expected for a passive coupling between the system and bath. However, in the context of the repeated interaction this non-standard thermalization is the expected result. Intuitively, in each iteration, the thermal qubits bring only their own on-site energy and, because $H_0+H_{B}$ is conserved, the interaction energy stored in the three-qubit system must be shared among six-qubits for a lapse of time $\tau$, decreasing the amount contained in the three-qubit system once the qubits from the bath are decoupled. After many iterations, the interaction energy in the system is depleted. Thus, in the steady state one expects $\langle H_{\rm int}\rangle=0$. 
According to this picture, the equilibrium state $\omega_\beta(H_0)$ is a natural consequence of the repeated coupling and uncoupling with the bath. 

\section{Conclusions} \label{sec: conclusions}

In this work, we have studied the non-equilibrium thermodynamics of an arbitrary quantum system in contact with many baths in the repeated interaction scheme. We have shown under which conditions the thermodynamic depends only on system operators and how this is carried on in the Lindblad limit obtaining the condition of quantum local detailed balance with respect to the equilibrium state $\omega_{\beta_r}(H_0)$ reached by the system when it is in contact just with the bath $r$. When $H_0=H_S$ this is the quantum local detailed balance of \cite{Spohn LDB}.

We have focused on the Lindblad equation of the three-qubit refrigerator to show that it can be microscopically derived under the repeated interaction scheme for any value of the interaction strength $g$ and to obtain the thermodynamic quantities associated to the process that it describes. Since the three-qubit refrigerator satisfies the quantum local detailed balance condition with respect to $\omega_{\beta_r}(H_0)$ with $H_0$ in Eq.(\ref{ecc: free Hamiltonian refri}), all thermodynamic quantities can be computed from the sole knowledge of the Lindblad equation without reference to the details of the bath.
Importantly, even though there is no explicit time dependence in the system Hamiltonian or dissipator, external power is required to bring the system towards its steady state. The amount of power scale with $g$ and therefore it is negligible when $g\to 0$. 
 
%This means that if one does not specify how the Lindblad equation is obtained, different microscopic approaches can lead to different thermodynamic quantities, thus, no statement on the thermodyanmics should be made. This would be the case, for example, if there exist an alternative approach (e.g. weak-coupling) to derive the Lindblad equation of the refrigerator, Eq.(\ref{ecc: Lindblad equation refrigerator}).

One may ask whether from another microscopic approach one would derive the same Lindblad equation but with different thermodynamic quantities such that no work will be needed to
bring the system to its steady state thus having a full autonomous three-qubit absorption refrigerator. We argue that for systems with quantum local detailed balance, because the thermodynamics is local (no reference to the environment is needed), it is the only appropriate thermodynamics for the model introduced in \cite{linden,linden2} for non-negligible values of $g$. 

On the other hand, for systems that do not obey the quantum local detailed balance, or systems described by maps without equilibrium \cite{StochPRE}, a description of the environment is necessary for the thermodynamic description.

\section*{Acknowledgements}

F.B. gratefully acknowledges the financial support of FONDECYT grant 1151390. C.L. gratefully acknowledges the financial support of CONICYT through Becas Magister Nacional 2016, Contract No. 22161809.

\end{document}